\documentclass[twoside]{article}
\usepackage{fleqn,espcrc2,epsf}

\title{QCDOC: A 10-teraflops scale computer for lattice QCD
\thanks{Talk presented at Lattice 2000, Bangalore, India.  This 
                research was supported in part by the U.~S.~Department of 
                Energy and the RIKEN-BNL Research Center.}}

\author{
D.~Chen\address[IBM]{IBM T. J. Watson Research Center, Yorktown Heights, NY, 10598},
N.~H.~Christ\address[Columbia]{Department of Physics, Columbia 
University, New York, NY, 10027}, 
C.~Cristian\addressmark[Columbia], 
Z.~Dong\addressmark[Columbia],
A. Gara\addressmark[IBM],
K.~Garg\addressmark[Columbia], 
B.~Joo\address{Department of Physics, University of Kentucky, Lexington, KY, 40506},
C.~Kim\addressmark[Columbia], 
L.~Levkova\addressmark[Columbia], 
X.~Liao\addressmark[Columbia],
R.~D.~Mawhinney\addressmark[Columbia],
S.~Ohta\address{Institute for Particle and Nuclear Studies,
KEK, Tsukuba, Ibaraki, 305-0801, Japan}\address[RBRC]{RIKEN-BNL 
Research Center, Brookhaven National Laboratory, Upton, NY, 11973},
T.~Wettig\addressmark[RBRC]\address{Department of Physics, Yale University,
New Haven, CT, 06520-8120} }
     
\begin{document}

\begin{abstract}
The architecture of a new class of computers, optimized for lattice QCD
calculations, is described.  An individual node is based on a single 
integrated circuit containing a PowerPC 32-bit integer processor with 
a 1 Gflops 64-bit IEEE floating point unit, 4 Mbyte of memory, 8 Gbit/sec 
nearest-neighbor communications and additional control and diagnostic 
circuitry.  The machine's name, QCDOC, derives from ``QCD On a Chip''.
\vspace{1pc}
\end{abstract}

\maketitle

\section{INTRODUCTION}

The numerical evaluation of Euclidean-space Feynman path integrals
provides a unique and powerful tool to study non-perturbative
phenomena in quantum field theory.  These techniques permit both
qualitative and quantitative study of low-energy hadronic physics
through first-principles, Quantum Chromodynamics calculations.
These methods also hold the promise of revealing new non-perturbative
phenomena that may be present in other quantum field theories 
that are potential candidates for the theory beyond the standard 
model.

Unfortunately the corresponding calculations are very demanding,
requiring large resources and sophisticated algorithms.  While a 
fully physical simulation including the effects of light quarks with
their physical masses is probably more than a decade away, there is
much optimism that physical results can be obtained by careful
extrapolation from parameter ranges which are less demanding
computationally.  Never-the-less, continued progress in this important
area of theoretical physics requires significant advances in
computational methods and active exploitation of the rapid progress
in microelectronics and computing technology.

Since the fundamental physics of low energy relativistic quantum 
field theory is accurately captured by the present lattice gauge 
theory formulation, it is appropriate to employ the largest possible 
computer resources to address outstanding problems.  In particular, 
much progress has been made over the past two decades by using specially 
designed computers, optimized to the particular characteristics of 
lattice QCD calculations\cite{Sexton:1995zz,Christ:1999ax}.  A massively 
parallel computer with a large number of computational nodes, a 
relatively small memory per node and relatively modest disk 
bandwidth and storage capacity per node is usually appropriate.  
However, relatively fast, low-latency inter-processor communication 
is often needed.  As a rough guide, for a fixed processor speed one
might require a processor-memory bandwidth (in words/sec) that is 
roughly one third of the processor speed (in floating point operations/sec).  
The total off-node bandwidth (counting both incoming and outgoing data), 
specified in words/second, should be roughly one tenth of this 
processor speed\cite{Aoki:1991vc}.

One example of such optimized computer construction is provided by
the present set of QCDSP machines\cite{Chen:1998cg,Mawhinney:2000fx}.  
Designed and constructed during the period 1993-1998 by the group centered 
at Columbia, these ``QCD on Digital Signal Processor'' machines are now 
installed and operational at Columbia University (400 Gflops), the 
RIKEN Brookhaven Research Center (600 Gflops) and the Thomas Jefferson 
Laboratory (50 Glfops).  By providing only the computer resources
required for lattice QCD, these machines achieve a favorable
cost performance figure of \$10/Mflops.

We have now begun the design of a new class of parallel machines
which represent further evolution of the architecture of the
QCDSP machines.  In the following we will describe our present
plans for these new machines.  After a brief discussion of the
QCDSP machines (Section~\ref{sec:QCDSP_review}), we will discuss 
the overall architecture of the new computer 
(Section~\ref{sec:architecture}), the features of 
the somewhat complex integrated circuit that lies at its core 
(Section~\ref{sec:ASIC}), the properties of the PowerPC RISC processor 
that will perform the actual computation (Section~\ref{sec:PowerPC}), 
our network/communications strategy for interprocessor communication
(Section~\ref{sec:network}) and a little about the software environment 
that we are planning (Section~\ref{sec:software}).

\section{QCDSP MACHINES}
\label{sec:QCDSP_review}

The present machines running at Columbia, the RIKEN-BNL Research
Center and Jefferson Laboratory are configured as four-dimensional arrays
of processing nodes, in quantities of 8192, 12,888 and 1024 nodes 
respectively.  Each node is made up of a Texas Instruments, TMS320C31-50 
digital signal processor, 2 Mbytes of DRAM (with an additional 0.5 Mbytes
of redundancy for error detection and correction), and an $\approx 250$K
transistor, application specific integrated circuit (or ASIC) which provides
a buffered/prefetching interface to the memory and eight 50 MHz serial
communication ports.  

\begin{figure}[t]
\vskip 0.1in
\epsfxsize=2.9in
\epsfbox{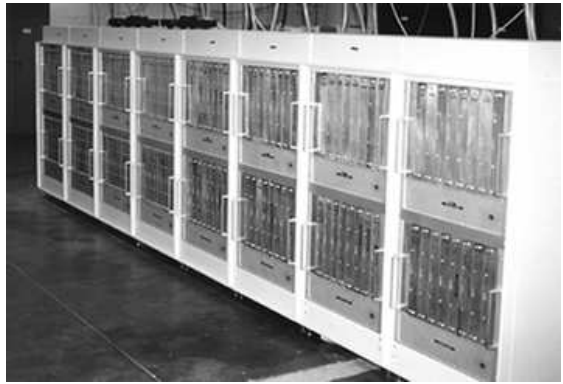}
\vskip -.4in
\caption{The 8,192-node, 0.4Tflops peak speed, {\it
QCDSP} machine running at Columbia since 4/98.}
\label{fig:128mb}
\vskip -0.3in
\end{figure}

Each node is mounted on a small daughter board.  Sixty-four such 
nodes are mounted on a mother 
board and eight mother boards fit into a backplane.  The 8,192-node machine 
at Columbia has 8 racks holding 16 backplanes and 128 mother boards and 
is shown in Figure~\ref{fig:128mb}.  

The four-dimensional, inter-node communication mesh is realized in the
following fashion.  First, the 64 nodes on each mother board are 
interconnected as a $4 \times 4 \times 2 \times 2$ hypercubic lattice.  Two 
of the $4 \times 2 \times 2$ faces, orthogonal to a common direction, are joined 
together in that direction creating a four-dimensional cylinder with six, 
three-dimensional faces corresponding to the remaining six of the eight faces 
of the original $4 \times 4 \times 2 \times 2$ hypercubic lattice.  Each of these 
six faces is connected to a separate cable brought out from the backplane.
These cables, six per mother board, can then be interconnected to create 
the desired overall machine topology, including a disconnected
collection of independent machines.  For example, the 12,288-node machine
at Brookhaven is currently operating as one 4,096-node machine and four
2,048-node machines.  This ability to cable the machine on the mother board
level provides valuable flexibility but also some inconvenience
when the cables must be manually rearranged.

\section{QCDOC ARCHITECTURE}
\label{sec:architecture}

With this background discussion of the current QCDSP machines, we now
turn to a general description of the architecture of the next QCDOC
computers.  For this next-generation machine we have followed a similar
strategy.  We seek to combine a large number of inexpensive, small, 
low-power processors into a machine capable of applying their 
computational power to a single very difficult calculation.   In this
way we attempt to optimize both the cost performance and operating
costs of the machine without compromising our ability to focus 
very significant computer resources on the most demanding problems.

Recall that the difficulty of a full QCD lattice calculation scales 
as a very high power of the volume: ${\rm Work} \sim L^{8-10}$, where 
$L$ is the linear lattice size.  As the problem gets larger the amount 
of computing power needed per volume increases rapidly, forcing us 
in the direction of many processors, each managing a decreasing 
fraction of the total physical volume.  

The network bandwidth and latency are therefore chosen to permit a
single problem to be mounted on a large machine.  However, we also
attempt to achieve sufficient flexibility that a small version of the
machine can do interesting physics as well and that a large machine
can be easily subdivided to tackle independent problems that may 
represent too small a lattice to require or fit on the full machine.

A critical part of the present design grows from our collaboration
with IBM and the resulting ability to exploit state-of-the-art IBM 
technology.  Using the next generation of IBM's ASIC technology, we
are designing a single integrated circuit, which will integrate the
complete functionality of our previous daughter board and nearly all
the circuitry of the mother board as well.  This follows the industry
trend of exploiting the decreasing semiconductor feature size to
build a ``system on a chip''.

In our case we will be able to incorporate an industry standard RISC
integer processor; a fully integrated 1 Gflops, 64-bit floating point 
auxilary processor; 4 Mbytes of DRAM; all inter-node communication 
and an Ethernet controller for external disk I/O as well as diagnostic 
and boot-up purposes, all on a single chip.  We expect such a chip 
will consume 1-2 Watts, will occupy a die approximately 1 cm on a side 
and will permit an aggregate cost/performance figure of less than 
\$1/Mflops.

In order to provide greater flexibility in memory size per node, allowing
even a quite small machine to have interesting physics applications,
we will provide an industry standard, double-data-rate, synchronous
dram module for each node.  This will permit a commercial memory card
to be added, providing an additional 32 Mbytes to 0.5 Gbytes 
per node as required by physics goals and economic limitations.

As in the previous machine, we have adopted a mesh, nearest-neighbor
communication scheme.  This eliminates the need for a switch, a component
that can easily represent a signficant fraction of the cost of a large
machine with a fast but more general network.  As is described below,
we presently plan a network of dimension higher than four.  Even two
extra dimensions provide considerable flexibility in joining the machine
into a variety of disconnected four-dimensional hyperplanes, thereby
significantly reducing the need for the somewhat inconvenient recabling
required by the QCDSP design.

The last element of the QCDOC architecture to address is the general-purpose
network used to boot the machine, load code, extract results and provide
access to mass storage.  In our present QCDSP machines these capabilites 
are provided by a tree made up of SCSI links with the final connections
on each mother board realized using a TI serial protocol.  We plan to 
exploit the tremendous commercial developments in Ethernet devices to
replace this SCSI network with Ethernet.  The ASIC in each node will contain
a standard 100 Mbit/sec Ethernet controller allowing each node to be addressed
individually and interrogated by the host computer through a tree of commerical
Ethernet switches.  

At present we plan to join the Ethernet connections for each group of 
four nodes into an on-board Ethernet switch.  Each of these 16 Ethernet
switches will have a 100 Mbit/sec, off-board Ethernet connection through 
an external connector.  This will reduce the effective simulataneous
bandwidth available per node to $\approx 3$ Mbytes/sec.  This next layer of
100 Mbit/sec connections will then be joined into 1 Gbit/sec Ethernet connections
using external commercial hardware and with no further loss of bandwidth.  
Connecting multiple RAID disks to the resulting multiple, 1 Gbit/s Ethernet 
wires should allow full support for this 3 Mbyte/sec/node bandwidth 
giving an 8K-node machine an aggregate 24 Gbyte/sec bandwidth to disk.

Thus, from the view of the host computer an 8K-node QCDOC machine looks 
like a large Ethernet appliance with 8K distinct Ethernet addresses.  Since
this Ethernet provides the only control link to this machine, we must
provide an Ethernet ``reset'' capability.  This requires a further simple,
hardwired Ethernet interface which is independent of the PowerPC and the 
more complex Ethernet controller that the PowerPC must initialize before 
it can be used.  However, such a capability is also needed for other 
applications and a very attractive solution appears to have been already 
developed within IBM Research.

\section{ASIC DESIGN}
\label{sec:ASIC}

\begin{figure*}[ht]
\epsfxsize=5.90in
\hskip 0.2in
\epsfbox{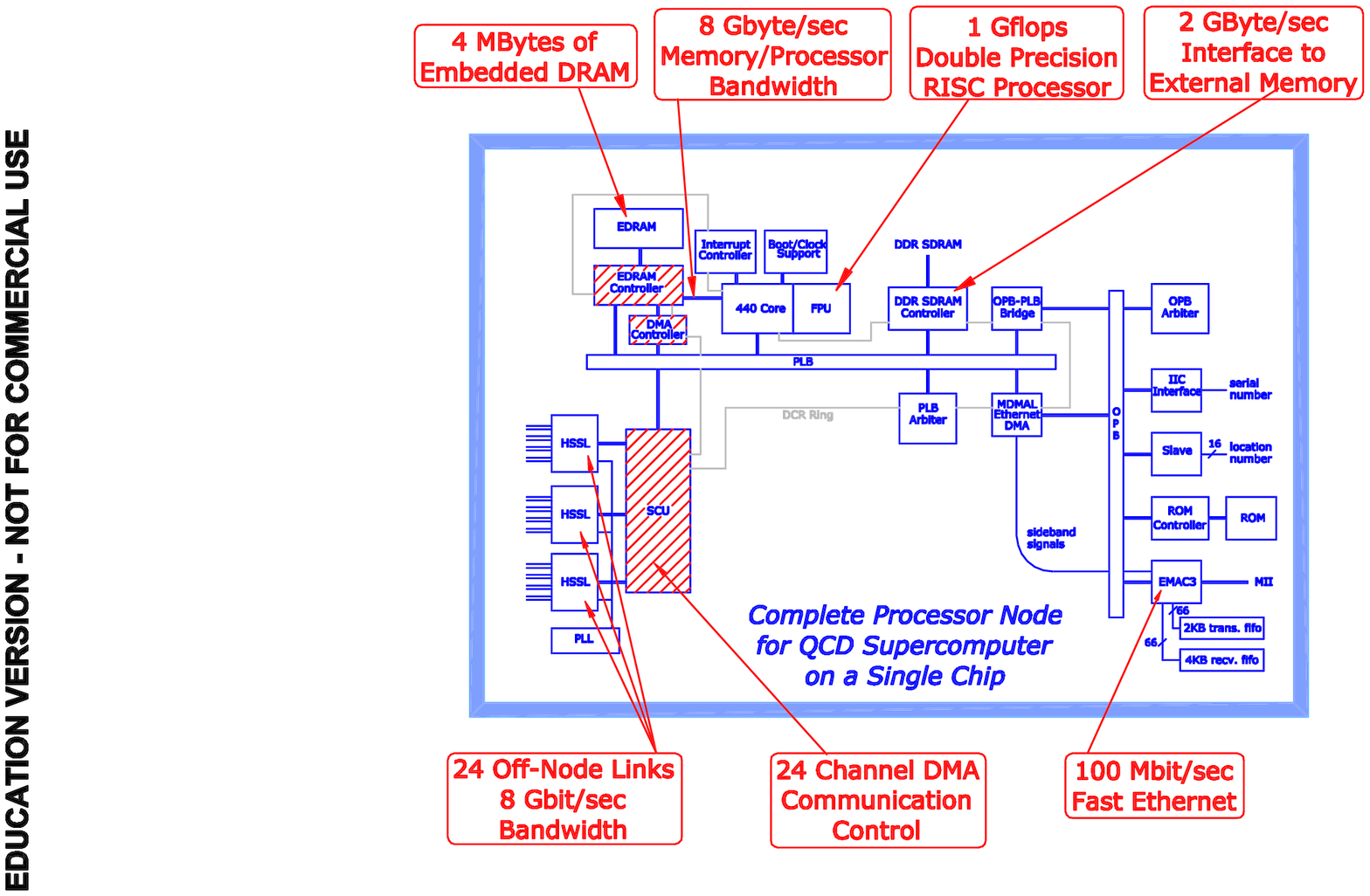}
\vskip -0.3in
\caption{Block diagram of the QCDOC ASIC design.  The cross-hatched
components are of our design while the remaining boxes represent
functions that are available as part of the IBM ASIC library.}
\label{fig:asic}
\vskip -0.1in
\end{figure*}
We next discuss the overall design of the application specific integrated 
circuit which, except for the external memory module, forms the entirety
of the new processing node.  This is best understood from Figure~\ref{fig:asic}.
The cross-hatched areas in the figure represent internal parts of the ASIC that
we must design while the open boxes are modules that are available as
library components that can be simply referred to in the hardware description
language version of the design.  (Including these pre-designed macros
is much like introducing a subroutine call into a normal computer program.)
A brief description of the various parts of design outlined in 
Figure~\ref{fig:asic} includes:

\subsection{PowerPC core}
This IBM-supplied macro represents the complete RISC processor
with its attached 1 Gflops, 64-bit IEEE floating point unit.
This is a model `440' PowerPC---a member of IBM's family of embedded 
PowerPC designs and is described below in Section~\ref{sec:PowerPC}.
A complete, functional model of the integer unit is represented 
in our simulation environment allowing us to execute compiled code 
on that portion of the ASIC as we begin the detailed design.

\subsection{Serial Communications}
This is provided by the serial communications unit (SCU) described below 
in Section~\ref{sec:network} and the three high speed serial modules,
labeled HSSL, in Figure~\ref{fig:asic}.  Each of these modules 
contains four independent sending ports and four independent receiving
ports, all operating at 500 MHz.  Each of the four serial receiving
ports collects incoming serial data into 8-bit units and provides
them to the SCU as bytes at 62.5 MHz.  Such high-speed components are
quite sophiticated, with built-in phase locking and a predetermined 
physical layout.  These three HSSL units, providing a total of 24 sending
or receiving ports, represent very valuable pre-packaged technology
that is supplied as part of the IBM ASIC design system.  When employed in
the geometry of a four dimensional mesh, only 8 of these links will be
used in each direction, providing a total off-node communications 
bandwidth of 8 Gbits/sec.

\subsection{EDRAM}
The 4 Mbytes of embedded DRAM provide sufficient storage that the data
for most lattice QCD problems can easily fit entirely within this memory.
Since we do not need to connect the memory and processor using external
drivers and pins, we can provide a much wider output bus from the memory.
In our design the memory controller is connected to the memory though
a 1024-bit bus (not including the bits needed for error correction and
detection).  This data is then carefully buffered into the 256-bit units
needed for cache line fetches and provided to the 440 core in 128-bit
units at 500 MHz.  Sufficient internal buffering is provided so that
sequential access can proceed at this 8 Gbytes/sec rate, hiding the DRAM
page misses that will necessarily occur as one moves through memory.

\subsection{External Memory controller}
An important IBM library component is the DDR SDRAM controller.  This unit
connects to the 128-bit Processor Local Bus (PLB), the standard, on-chip 
bus that also joins the PowerPC processor and the SCU.  This controller manages 
all aspects of external memory accesses including DRAM refresh and error 
detection and correction.  Both the PLB and the external memory 
will operate at 1/3 of the processor speed.  While the connection to the
external memory is only 72 bits (including error detection and correction),
the double data rate feature means that data is effectively clocked 
at twice the 166 MHz PLB bus frequency, giving a 2.6 Gbytes/sec bandwidth
to external memory.

\subsection{Ethernet Controller}
The final module described is the Ethernet controller.  This is a highly 
functional, pre-designed unit which will manage Ethernet traffic with 
infrequent interruption of the processor.  It is supplied with a direct 
memory access (DMA) unit and should also be supported by a pre-existing 
software driver.  This Ethernet controller is connected to the PLB 
somewhat indirectly through a second, 32-bit On-chip Perpherial Bus, 
again a standard bus within the IBM library of ASIC components.

\section{PowerPC PROPERTIES}
\label{sec:PowerPC}
The processor core, central to our design, is an industry standard, embedded
PowerPC RISC processor.  This is a 32-bit processor with 32 general purpose
registers, a 32 KByte data cache and a 32 KByte, prefetching instruction
cache.  The CPU can issue two instructions on every cycle, contains three
execution pipes, carries out branch prediction and supports out-of-order
instruction issue, execution and completion.  It supports highly functional
memory management connecting 32-bit effective and 36-bit physical addresses
using a 64-entry translation look-aside buffer, where each entry identifies
an independently mapped page of length between 1 Kbyte and 256 Mbyte.
The 64-bit IEEE floating point unit is connected as an auxilliary processor 
which executes Book-E floating point instructions in hardware with direct 
access to the processors data cache.

\section{COMMUNICATIONS/NETWORK}
\label{sec:network}
The communications network is a natural evolution of that used successfully
in the QCDSP machines.  The basic transfer size is increased from 32- to 
64-bits.  The inter-node communication is self-synchronizing with the receipt 
of a given 64-bit word acknowledged only after that word has been removed 
from the input buffer, indicating that another word can be sent without 
the possibility of data loss.  The detection of an error will cause the issue 
of an ``acknowledgement with error'' which will initiate a retry.  The 
communications protocol is designed so that any single bit error within 
32-bits will be detected.  If that error occurs during the first 8 bits 
of a transfer, those bits used to identify the transfer, the error will 
in addition be corrected allowing the proper response to the error to 
be taken.

In order that no communications bandwidth is lost waiting for an 
acknowlegement, four distinct receive buffers are provided with each 
separately acknowledged.  This permits four words to be sent before an 
acknowledgement is received.  These receive buffers will be divided into 
two groups.  The first group of three is used for normal data transfers 
with both the sent and received data streamed to memory by an independent 
DMA unit for each of the 24 external wires.  These data transfers will be
programmed as a sequence of block-strided moves controlled by simple
chained instructions loaded into the SCU.  The second group is the fourth
of these registers.  It is loaded and unloaded directly by the PowerPC
and data arriving is signaled by a processor interrupt.  This supervisor
communication channel can be used to support efficient operating system 
communication between neighboring nodes that is independent of ongoing 
application data transfers.

We expect that the topology of the communications network that is actually  
used will be the standard four-dimensional torus appropriate for Euclidean
Feynman path integration.   However, by providing a higher dimensional mesh
we will facilitate the subdivision of the machine in software reducing
the need to physically reconnect the communications cables when a different
set of physics jobs is to be run.

\begin{figure}[h]
\vskip -0.4in
\epsfxsize=2.9in
\epsfbox{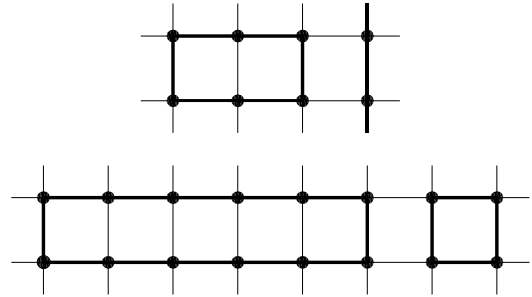}
\vskip -0.6in
\hskip -0.35in
\caption{Two examples of reducing a 2 dimensional torus
to a sum of one-dimensional tori.  Note, the wires leaving
the edges of each figure will be joined back to the other 
side of that figure.}
\label{fig:network}
\end{figure}
\vskip -0.3in

This ``reconfiguration through dimensional reduction'' can be most 
easily understood by examining some lower dimensional examples.
First consider what we would like ultimately to be a one-dimensional
machine of eight nodes.  If these nodes are interconnected into a 
two-dimensional, $4 \times 2$ mesh, we can realize a number
of different one-dimensional mesh configurations as shown in the upper
portion of Figure~\ref{fig:network}.   The darkened links shown in that 
Figure demonstrate a choice in which the 8-node machine is configured into two
partitions: a 6-node machine and a separate 2-node machine.  Clearly
a variety of other choices are possible as well including an 8-node
machine and two 4-node machines.

\begin{figure}[h]
\vskip -0.38in
\epsfxsize=2.9in
\epsfbox{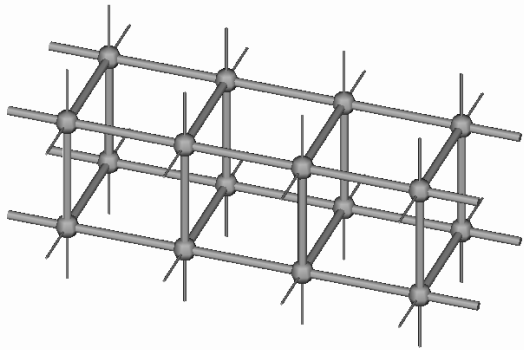}
\hskip -0.2in
\vskip -0.6in
\caption{Here the thicker connections represent a $4\times 4$, 
two-dimensional torus created from a $2 \times 2 \times 4$, 
three-dimensional mesh.  This mesh is also connected as a torus by joining the 
corresponding wires leaving opposite faces of the $2 \times 2 \times 4$
cube.}
\label{fig:network3}
\end{figure}

A more complicated example is shown in Figure~\ref{fig:network3}
where what might have been a simple $4\times4$, two-dimensional mesh machine
is instead wired as a three-dimensional $2\times 2 \times 4$ device.  As shown
in that figure, the original $4\times 4$ geometry is easily realized.
However, it is not difficult to recognize a $2 \times 8$ mesh or two 
$2 \times 4$ machines.

\vskip -0.25in

In order to see how this is accomplished for the six-dimensional case of
interest, it is easiest to consider an 
example.  As an illustration, consider an 8,192-node machine composed of
128 mother boards, each with 64 nodes, a likely machine configuration.
Further, we will interconnect these 64 nodes as a $2^6$ cube with three
pairs of faces joined back on themselves to realize a three-dimensional
torus on the mother board.  In an arbitrary set of coordinates, 
let us identify a node with 6 coordinates: $(n_0, n_1, ..., n_5)$.  
We might then choose the first three coordinates as corresponding to this
three-dimensional torus.  Thus, $n_i \in [0,1]$ for $i=0,\,1$ and 2.  Six of
the twelve faces of this $2^6$ cube have been connected to each other.
This leaves a final six faces (each of size $2^5=32$) to be connected to 
other mother boards through edge connectors on the mother board.  The 
required 192 signals (or 768 wires) is large but possibly managable.

For the next coordinate, $n_3$, we might connect together 4 mother 
boards within a single backplane and use $n_3 = k+2*m$, where $k \in [0,1]$
determines the third coordinate of the node within the $2^6$ hypercube on 
the mother board and $m \in [0,3]$, labels the mother board on which that 
node resides.  This group of 256 nodes is now a $2^3 \times 8 \times 2^2$
six-dimensional solid with 8 of its 12 faces joined to themselves.
This leaves two remaining directions to be joined, connecting this
group of 256 nodes with the remaining 32 groups within our example machine.
Each of the four faces of such groups of 256-nodes must be connected through
a separate group of cables to the neighboring face of another 256-node
group.  Since each of these faces is made up of $2^4 \times 8$ processors a 
total of 128 signals are required per face.  An eight mother board backplane 
would then need to provide connectors for $2 \times 4 = 8$ such groups of 
128 signals.  The total of 1024 signals is less than the 1,280 signals 
leaving the backplanes of our present QCDSP machines.

If we arrange these 256-node groups as a final $4 \times 8$ mesh, the 
final machine becomes a six-dimensional $2^3 \times 8^2 \times 16$ torus.  
While there may be computational problems for which this machine could 
be employed directly as a six-dimensional torus, we expect that the typical configuration would exploit the six-dimensional interconnect to realize 
a four-dimensional torus more appropriate for lattice gauge theory 
calculations.  One simple way to achieve such a reduction from six to 
four dimensions takes two independent two-dimensional factors and uses 
only a one-dimensional subgrid (or collection of one-dimensional subgrids) 
in each factor to produce a four-dimensional product.

For example, we can use the scheme in the lower diagram in Figure~\ref{fig:network}
twice to separate two of the $2 \times 8$ factors in the machine, each 
into two, one-dimensional terms, one of 12 nodes and one of 4 nodes. 
We can thereby partition our 8,192-node, $2^3 \times 8^2 \times 16$ machine, 
into four independent, four-dimensional tori: 
one $2 \times 12 \times 12 \times 16$, 
two $2 \times  4 \times 12 \times 16$ and
one $2 \times  4 \times 4 \times 16$.
This would permit a $24^3 \times 32$ calculation
to be done on more than one half of the hardware.  Since a reasonable programing
model requires that an even number of lattice sites appear on each node for
each dimension of the machine, fitting factors of three into our lattice  
is a non-trival accomplishent for a large machine which is even in each 
dimension. 

\section{SOFTWARE}
\label{sec:software}

We plan a software environment for this next generation of
computers which is a natural evolution of that available on
the present QCDSP machines.  This follows a ``data parallel''
programming model in which application code is written so that a 
single program runs on each node, executing essentially the same 
instructions with different data on each node.  The exceptions
to this pattern are usually I/O or communications routines
where the placement of disks or the pattern of communications
is not homogenous and requires different actions from different
processors.  The code is cross-compiled on a UNIX-based host
(to date this is always a SUN machine) and then down-loaded to 
the parallel machine.

A particular partition of the machine is controlled from an
extended UNIX shell environment which includes additional
commands allowing programs to be loaded and executed, data
to be loaded or read and individual memory locations to be examined. 
Both interactive and batch UNIX processes can be run within this 
environment.  This functionality is supported by
a further suite of operating system code that is executing
on the individual nodes.

This operating environment also provides `C'-like subroutines 
that can be called by application programs allowing {\tt printf()},
{\tt fopen()}, {\tt fclose()} and {\tt fprintf()} capability.
Critical to this software environment is carefully designed 
low-level code with a high degree of robustness and diagnostic 
capability permitting a hardware fault to be isolated and 
identified from software.  This underlying boot/diagnostic
kernel is essential to the maintenance of a system of 
more than 20K nodes.

The industry standard RISC processor will allow further
improvements on this reasonably convenient scheme.  Since the
Book-E compliant processor is supported by a number of
standard compilers we should be able to provide a well-supported
and highly functional C/C++ programming environment.  This is in
welcome constrast to the somewhat limited capabilities of the
C++ compiler available for the digital signal processors in the
present QCDSP machines.  An equally important enhancement results
from the highly functional memory management unit in the PowerPC
processor.  We plan to use this capability to isolate system and
application code, creating a reasonably robust code 
development/debugging environment.

Of course, with such a flexible processor, an even more 
sophisticated software environment is certainly possible.  While 
LINUX on every node could certainly be provided, this degree of 
generality may well be inconsistent with high performance for 
QCD applications.

\section{CONCLUSION}
This next-generation, QCDOC architecture described above will provide 
a very significant advance over our present QCDSP machines.  We
anticipate a cost performance of better than \$1/Mflops, a $10\times$
improvement on the QCDSP machines.  Given the large processor/memory
bandwidth, optimized QCD code should sustain above 50\% on the 
new machine and even generic `C' code should execute with reasonable
efficiency.  We plan large machines at Columbia, the RIKEN
Brookhaven Research Center, a UKQCD machine in Edinburgh and a possible
national machine for the US lattice QCD community.

\end{document}